\documentclass[floatfix,amsmath,amsfonts,twocolumn]{revtex4}  
\usepackage{enumitem}
\usepackage{amsmath}
\usepackage{graphicx}
\usepackage{epsfig}
\usepackage{bm}
\usepackage{bbold}
\usepackage{amssymb}
\usepackage{color}

\newcommand{\PT}{{\cal PT}}

\begin{document}

\title{Nonlinear Schr\"odinger equations with  amplitude-dependent Wadati potentials}

\author{Dmitry A. Zezyulin\footnote{email: d.zezyulin@gmail.com}}

\affiliation{Department of Physics and Engineering, ITMO University, St.~Petersburg 197101, Russia}

\date{\today}

\begin{abstract}
Complex Wadati-type potentials of the form $V(x)=-w^2(x) + iw_x(x)$, where $w(x)$ is a real-valued function, are known to possess a number of intriguing features, unusual for generic non-Hermitian potentials. In the present work, we introduce a class of nonlinear Schr\"odinger-type problems  which generalize the Wadati potentials by assuming that the base function $w(x)$ depends not only on the transverse spatial coordinate but also on the amplitude of the field.   Several examples of prospective  physical relevance are discussed,   including    models with the nonlinear dispersion or with the derivative nonlinearity. The numerical study indicates  that the generalized   model inherits the remarkable features of standard  Wadati potentials, such as the existence of   continuous soliton families, the possibility of symmetry-breaking bifurcations   when the model obeys the parity-time symmetry, the existence of constant-amplitude waves, and the eigenvalue quartets in the linear-instability spectra. Our results deepen the current understanding of the interplay between nonlinearity and non-Hermiticity and expand the class of  systems which enjoy the exceptional combination of properties   unusual for generic dissipative    nonlinear models.
\end{abstract}

\maketitle

\section{Introduction}

The combination of  nonlinearity and non-Hermiticity can  have a dramatic impact on the properties of a waveguiding system.  In particular, the presence of a  non-Hermitian complex potential  (which in physical terms takes into account the energy exchange with the environment) can 
heavily alter the structure of stationary localized nonlinear modes or solitons propagating in the system. While in   conservative waveguides the solitons exist as continuous families parameterized by an `internal' parameter, such as the soliton frequency or amplitude, the dissipative solitons most usually exist  as isolated points \cite{AA,Rosanov} which, from the dynamical point of view, behave as attractors (provided that the soliton is stable). A prototypical model that  illustrates this dissimilarity between conservative and dissipative systems is   the generalized nonlinear Schr\"odinger equation (GNLSE)   
\begin{eqnarray}
\label{eq:intro}
i\Psi_t = -\Psi_{xx} + V(x)\Psi   + W(x) F(|\Psi|^2) \Psi,
\end{eqnarray}
where complex-valued functions $V(x)$ and $W(x)$ can be referred to as a linear and nonlinear potential, respectively, and  real-valued function $F(\cdot)$, such that $F(0)=0$, specifies the   nonlinearity. Equation (\ref{eq:intro}) is a canonical model describing evolution of nonlinear waves in various physical settings. In particular, in optical applications, $\Psi$ corresponds to the dimensionless amplitude of the electric field. Its dependence on coordinate $t$ describes  evolution of the pulse along the propagation direction, and   $x$ is the transverse coordinate. Effective complex optical potentials correspond to the presence of spatial regions with gain and loss \cite{KYZ,Ganainy}, which, in particular, can be implemented using coherent multilevel atoms driven by external laser fields \cite{HH17,HHWadati}.

When the imaginary parts of both potentials vanish, the model becomes conservative, and its stationary nonlinear modes $\Psi(x,t) = e^{-i\mu t}\psi(x)$  can be   parameterized by the   continuous change of the real frequency $\mu$. However, if at least one of the potentials is complex, then    the continuous families of solitons  typically disappear. At the same time,   there exist at least two types of   non-Hermitian potentials that do support continuous families of nonlinear modes (see a more detailed discussion in  \cite{ZSA}). Systems of the first type obey the parity-time ($\PT$) symmetry \cite{Bender,KYZ}: in this case real and imaginary parts of functions $V(x)$ and $W(x)$ are even and odd functions of $x$, respectively \cite{Musslimani,AKKZ}. The second type corresponds to the so-called Wadati-type potentials \cite{Wadati} for which
\begin{equation}
\label{eq:Wadati}
V(x)=-w^2(x) + iw_x(x), \quad W(x)\equiv 1,
\end{equation}
where   $w(x)$ is a real-valued differentiable function which is not required to have any particular symmetry (at the same time, if $w(x)$ is even, then the  resulting Wadati potential $V(x)$ becomes $\PT$ symmetric, i.e., the two types of potentials are partially overlapping). In what follows, it will be convenient to say that the GNLSE (\ref{eq:intro}) with  potentials (\ref{eq:Wadati}) corresponds to \textit{linear} Wadati potentials.   Apart from the existence of continuous families of nonlinear modes \cite{Tsoy},   linear Wadati potentials are known to support several other remarkable features.   When the nonlinearity is absent, i.e.,    $F\equiv 0$, linear Wadati potentials can have all-real spectrum of eigenvalues \cite{NixonYang16PRA} and undergo   distinctive phase transition from all-real to complex spectrum through an exceptional point or a self-dual spectral singularity \cite{Yang17,ZK20}. Returning to the nonlinear setup,  Wadati-type potentials support  the existence of constant-amplitude nonlinear waves \cite{Makris}, feature unusual dynamical behavior  near the phase-transition threshold \cite{integral}, and   eigenvalue quartets in the linear-stability spectrum \cite{quartets}. Additionally, when a Wadati potential is $\PT$ symmetric, it allows for bifurcations of families of non-$\PT$-symmetric modes \cite{SB}, which is impossible for $\PT$-symmetric potentials of general form. 
While the   theory of Wadati potentials is far from being complete,  on the qualitative level some of their peculiar properties can be  explained by the existence of  a  `conserved' (i.e., $x$-independent) quantity which constrains the shape of  stationary modes \cite{KonZez14}.  

Unusual and not yet fully understood     properties of Wadati potentials   encourage to look for  their possible  generalizations. In the meantime,  most of the presently available studies of Wadati potentials  are  basically limited by    GNLSE with spatially-uniform power-law and saturating nonlinearities \cite{NixonYang16,saturable}.   In the present paper,   we introduce a more profound generalization of  Wadati potentials by considering the situation where  the base function $w(x)$ depends not only on the spatial coordinate $x$ but also on the amplitude of the field $\Psi$. We argue that the extended system preserves  some of the properties of   linear   Wadati potentials. Particular realizations of the introduced generalization  lead to nonlinear systems of potential physical relevance. Those    include a GNLSE equation with the  additional  higher-order nonlinearity that emerges due to the stimulated response in  optical fibers  and a GNLSE   with the  derivative nonlinearity.  Combining our analytical understandings with simple demonstrative computations,   we construct continuous families of nonlinear modes and discuss their linearization spectra. Additionally, we show that when the generalized system is $\PT$ symmetric, it undergoes a symmetry-breaking bifurcation which results in  continuous families of non-$\PT$-symmetric solitons.

The rest of the paper is organized as follows.  Section~\ref{sec:model} outlines the derivation of the generalized model  and presents several particular realizations of the  extended system. Section~\ref{sec:case} contains a case study of a model with the derivative nonlinearity. Section~\ref{sec:concl} concludes the paper.

\section{Construction of generalized potentials}
\label{sec:model}

For stationary modes   $\Psi(x,t) = e^{-i\mu t}\psi(x)$,  equation (\ref{eq:intro}) with linear Wadati potential (\ref{eq:Wadati})  becomes
\begin{equation}
\label{eq:old}
\mu \psi = -\psi_{xx} + (-w^2 + iw_x) \psi  + F(|\psi|^2) \psi.
\end{equation}
The latter stationary equation has been studied in the previous literature \cite{Tsoy,KonZez14,SB,NixonYang16,saturable,quartets,BarZezKon,Yang20}. 
In particular, it is known  that if Eq.~(\ref{eq:old}) is considered as a dynamical system with $x$ playing the role of the evolution variable, then the respective `dynamics' is constrained by a conserved (i.e., $x$-independent) quantity \cite{KonZez14}.  Our construction of the generalized model relies on a `gauge transformation' which converts the stationary    equation (\ref{eq:old})   to another   ordinary differential equation, where the   $x$-independent    `integral of motion'  has an especially simple form  \cite{unpublished}. Indeed, using the substitution $\psi(x) = \phi(x)e^{i\int w(x) dx}$,  where $\phi(x)$ is a new stationary field, from (\ref{eq:old}) we obtain
\begin{equation}
\label{eq:stat:w}
\mu \phi = -\phi_{xx}  - 2i w(x) \phi_x + F(|\phi|^2) \phi.
\end{equation}
Multiplying the latter equation by $\phi^*_x$ and adding it with its complex conjugate, we obtain the integral of motion 
\begin{equation}
\label{eq:integral}
\mu |\phi|^2  =  -  |\phi_x|^2 + \int_0^{|\phi|^2 } F(\xi) d\xi  +  \textrm{const},
\end{equation} 
where `const' is an arbitrary $x$-independent quantity. Clearly, for localized modes with $\lim_{x\to\pm\infty}\phi(x)=\lim_{x\to\pm\infty}\phi_x(x)=0$ this constant must be  zero.  
Using the obtained  integral of motion   as a qualitative argument, continuous families of nonlinear localized modes have been constructed in linear Wadati potentials \cite{KonZez14}.   In the meantime, since the integral (\ref{eq:integral}) does not contain function $w(x)$ explicitly,   Eq.~(\ref{eq:stat:w}) can be generalized naturally    by assuming that the base function $w(x)$ depends not only on the spatial coordinate $x$ but also  on the amplitude of the field.  This idea suggests to replace Eq.~(\ref{eq:stat:w}) with the following more general one:
\begin{eqnarray}
\mu \phi  =  -\phi_{xx}  - 2i A(x,|\phi|^2) \phi_x + F(|\phi|^2) \phi,
\label{eq:stat}
\end{eqnarray}
where $A(\cdot,\cdot)$  is a real-valued function of two variables. It is easy to check that the identity   (\ref{eq:integral}) remains valid for the newly introduced   Eq.~(\ref{eq:stat}).  
We further make the `inverse'   gauge transformation $\phi(x) =  \psi(x) e^{i\int z(x) dx}$, where $z(x):=B(x, |\psi|^2)$ is  another real-valued function,  which converts (\ref{eq:stat}) into the following equation:
\begin{widetext}
\begin{equation}
\label{eq:stat:psi}
\mu \psi = -\psi_{xx}  + \left (2AB+B^2 - i\frac{\partial B(x, |\psi|^2)}{\partial x} - i\frac{ \partial B(x, |\psi|^2)}{\partial |\psi|^2} (|\psi|^2)_x \right)\psi  - 2i(A+B)\psi_x +    F(|\psi|^2) \psi.
\end{equation}
The obtained stationary   equation corresponds to the  following temporal evolution   problem:
\begin{equation}
\label{eq:main}
i  \Psi_t = -\Psi_{xx}  + \left (2AB+B^2 - i\frac{\partial B(x, |\Psi|^2)}{\partial x} - i\frac{ \partial B(x, |\Psi|^2)}{\partial |\Psi|^2} (|\Psi|^2)_x \right)\Psi  - 2i(A+B)\Psi_x +    F(|\Psi|^2) \Psi.
\end{equation}
\end{widetext}
Equation (\ref{eq:main}) is the central point of the present paper. It represents the generalization of the previously studied GNLSE   with linear Wadati potentials, which can be recovered from   Eq.~(\ref{eq:main}) by setting   $A=-B = w(x)$. To distinguish this model from the known case of  linear Wadati potentials, we will say that Eq.~(\ref{eq:main}) corresponds to \textit{nonlinear} Wadati potentials. By construction, the stationary version of this model [i.e., Eq.~(\ref{eq:stat:psi})] has the $x$-independent conserved quantity and is therefore expected to support continuous families of nonlinear localized modes and  share   other properties of linear  Wadati potentials.

For different choices of $A$ and $B$  the obtained equation (\ref{eq:main}) acquires specific shapes which can be of potential relevance for physical applications.

\paragraph*{Example 1.} Choosing   $A = -B = w(x) + \sigma |\Psi|^2$, where $\sigma$ is a real coefficient, from (\ref{eq:main}) we obtain the following equation:
\begin{eqnarray}
i\Psi_t = -\Psi_{xx} + (-w^2 + i w_x)\Psi - 2\sigma w(x) |\Psi|^2\Psi \nonumber\\[2mm]- \sigma^2 |\Psi|^4\Psi + i\sigma (|\Psi|^2)_x \Psi +  F(|\Psi|^2) \Psi.
\label{eq:SRS}
\end{eqnarray} 
The case  $\sigma=0$   recovers the   previously considered GNLSE    with linear   Wadati potentials.  Let us comment on the newly appeared terms that contain the coefficient $\sigma$. The term  $ 2\sigma w(x) |\Psi|^2\Psi$ can be considered as   cubic nonlinearity with a real-valued nonlinear potential whose spatial shape is given by the   base function $w(x)$. The  term  $\sigma^2 |\Psi|^4\Psi$ corresponds to the spatially uniform  quintic focusing nonlinearity. Assuming that  $\sigma$ is small, this term can be neglected in the leading order. In fact  this term   can   be `eliminated' by  a simple redefinition of function $F$.   The most unconventional term in Eq.~(\ref{eq:SRS})   corresponds to  $i\sigma (|\Psi|^2)_x \Psi$. In fiber optics, the higher-order nonlinear terms of this form   have been used to take into account nonlinear dispersion resulting from the  stimulated Raman scattering (SRS) of ultrashort pulses, see e.g. \cite{SRS1,SRS2,SRS3,SRS4}.   

\paragraph*{Example 2.} Choosing $A = -B = \sigma w(x) |\Psi|^2$, we arrive at the following equation, where  the nonlinear gain-and-loss term and nonlinear   dispersion are spatially modulated by function $w(x)$ and its derivative $w_x(x)$:
\begin{eqnarray}
\label{eq:}
i\Psi_t = -\Psi_{xx}  + (-\sigma ^2 w^2 |\Psi|^4 + i \sigma   w_x |\Psi|^2)\Psi\nonumber\\[2mm]
   + i \sigma w(x) (|\Psi|^2)_x\Psi  +  F(|\Psi|^2) \Psi.
\end{eqnarray}

\paragraph*{Example 3.}  The choice  $A =  w(x)  - \sigma|\Psi|^2/2$ and $B=-w(x) + \sigma|\Psi|^2$ transforms equation (\ref{eq:main})  to the following model:
\begin{eqnarray}
i\Psi_t = -\Psi_{xx}  + (-w^2 + i w_x)\Psi + \sigma w(x) |\Psi|^2\Psi   \nonumber\\[2mm] +  F(|\Psi|^2) \Psi - i\sigma (|\Psi|^2\Psi)_x.
\label{eq:dnls}
\end{eqnarray}

In the latter equation,  the spatially inhomogeneous cubic nonlinearity  $ \sigma w(x) |\Psi|^2\Psi$ is present   together with the \emph{derivative} nonlinearity $i\sigma(|\Psi|^2\Psi)_x$. The derivative NLSE  $i\Psi_t = -\Psi_{xx} - i\sigma (|\Psi|^2\Psi)_x$  is   integrable by the inverse scattering method \cite{Kaup,CCL} and has been widely  used to describe  nonlinear waves in plasma \cite{Miolhus,Mio} and optical fibers \cite{Agrawal}. Notice that the `additional' terms in Eq.~(\ref{eq:dnls}) (i.e., those with coefficient $\sigma$) are conservative, i.e., do not contribute to the power balance equation: indeed,  introducing the power $P(t) = \int_{-\infty}^\infty |\Psi|^2 dx$, from Eq.~(\ref{eq:dnls}) we compute
\begin{equation}
\frac{dP(t)}{d t} = 2\int_{-\infty}^\infty w_x |\Psi|^2dx.
\end{equation}
 Several  recent studies have addressed the derivative NLSE   with     $\PT$-symmetric potentials  \cite{Saleh14,DNLS1,DNLS2,DNLS3}.  Being motivated, in part, by this newly emerged interest in non-Hermitian extensions of the derivative NLSE,   in the next section, we perform a more detailed examination  of the   model (\ref{eq:dnls}).

\section{Wadati potentials with the derivative nonlinearity}
\label{sec:case}

\subsection{Constant-amplitude waves}

In spite of their spatially inhomogeneous structure, the linear Wadati potentials are known to support constant-amplitude solutions \cite{Makris}. Similar solutions can be found in the introduced generalized models. They can be constructed easily by noticing that a pair $\phi(x)\equiv \rho_0$ and $\mu = F(\rho_0^2)$, where $\rho_0$ is arbitrary real amplitude,  solves Eq.~(\ref{eq:stat:w}). In particular, for Eq.~(\ref{eq:dnls})  the constant-amplitude solution reads  
\begin{equation}
\Psi(x,t) = \rho_0\exp  \left\{ i\int w(x)dx  - i\sigma \rho_0^2 x -i F(\rho_0^2) t\right \}.
\end{equation}
In linear Wadati potentials, constant-amplitude  waves   of the analogous   form have   been used to examine the development of modulational instability in   non-Hermitian optical media \cite{Makris} and  to implement coherent perfect absorption of nonlinear waves \cite{ZK20}.

\subsection{Computing the continuous families of  nonlinear modes using the numerical shooting approach}

Let us now use the demonstrative computation to argue that Eq.~(\ref{eq:dnls}) supports continuous families of stationary localized modes $\Psi = e^{-i\mu t} \psi(x)$, where $\mu$ is a real parameter, and $\lim_{x\to\pm \infty} \psi(x)= 0$. The corresponding procedure  is essentially a   generalization of the approach  from \cite{KonZez14}.  It is more convenient to describe it  in terms of the equivalent  Eq.~(\ref{eq:stat}).  We  reduce the dimensionality of the underlying phase space by employing   the representation $\phi(x) = \rho(x)\exp\{i\int v(x) dx\}$, where $\rho(x)$ and $v(x)$ are real-valued functions. Separating Eq.~(\ref{eq:stat}) into real and imaginary parts, we arrive at the system
\begin{eqnarray}
\label{eq:hydro1}
\mu \rho = -\rho_{xx} + v^2\rho + 2A(x,\rho^2) v \rho  + F(\rho^2)\rho,\\[1mm]
\label{eq:hydro2}
-v_x\rho-2v\rho_x - 2A(x,\rho^2)\rho_x=0.
\end{eqnarray}
We fix some   $\mu<0$ and   assume that $A(x, \rho^2(x))$ decays fast as $x\to\pm \infty$. Then solutions $\phi_\pm(x)$ that decay, respectively, at  $+\infty$  and $-\infty$   have   simple asymptotic behavior $\phi_\pm(x) = e^{-\sqrt{|\mu|}|x|}(C_\pm + o(1)_{x\to \pm \infty})$, where $C_\pm$ are real  $x$-independent constants. The   asymptotic behavior of derivatives $\phi_{x,\pm}(x)$ can be obtained by  differentiating  the asymptotic equations for  $\phi_\pm(x)$. Choosing some sufficiently large $x_\infty \gg 1$, for any  $C_+$ and $C_-$ we can use the found asymptotic behavior to approximate  $(\rho_\pm(\pm x_\infty),  \rho_{x,\pm}(\pm x_\infty), v_\pm(\pm x_\infty))$ and then use those   as initial values for numerical solution of system (\ref{eq:hydro1})--(\ref{eq:hydro2}), computing in this way  solutions $(\rho_+(x),  \rho_{x,+}(x), v_+(x))$ on the interval $[0, x_\infty)$ and     $(\rho_-(x),  \rho_{x,-}(x), v_-(x))$ on the interval $(-x_\infty, 0]$. Looking for a  continuously differentiable localized mode $\phi(x)$ which decays at both infinities, we need to find a solution  to the following  system of three equations:
\begin{equation}
\label{eq:3eqs}
\rho_+(0) =  \rho_-(0), \quad  \rho_{+,x}(0) =  \rho_{-,x}(0), \quad v_+(0)=v_-(0)
\end{equation}
with respect to two unknowns (i.e., two `shooting parameters') $C_+$ and $C_-$. At   first glance,   system (\ref{eq:3eqs}) seems overdetermined and therefore is not expected to have a  solution. However,  the integral (\ref{eq:integral}) imposes an additional relation between  the functions: $\mu \rho_\pm^2(x) = -   \rho_{\pm,x}^2(x) - v^2_\pm(x)\rho_\pm^2(x) + \int_0^{\rho_\pm^2(x)} F(\xi) d\xi$.  These constraints imply that if any two equations (say the first two) of system (\ref{eq:3eqs}) are satisfied, then the third equation is satisfied automatically  (speaking more precisely, the third equation will be satisfied in the form $v_+^2(0)=v_-^2(0)$; however, the spurious  solutions with $v_+(0)=-v_-(0)$ can be easily filtered out in the practical realization of the procedure).  Therefore,    system (\ref{eq:3eqs}) is not overdetermined and can be  expected to have one or several solutions each of which corresponds to a localized mode. Since the described procedure can be fulfilled for any $\mu<0$,   the continuous in $\mu$ families of localized modes are indeed  expected to exist.

\begin{figure}
	\begin{center}
		\includegraphics[width=0.99\columnwidth]{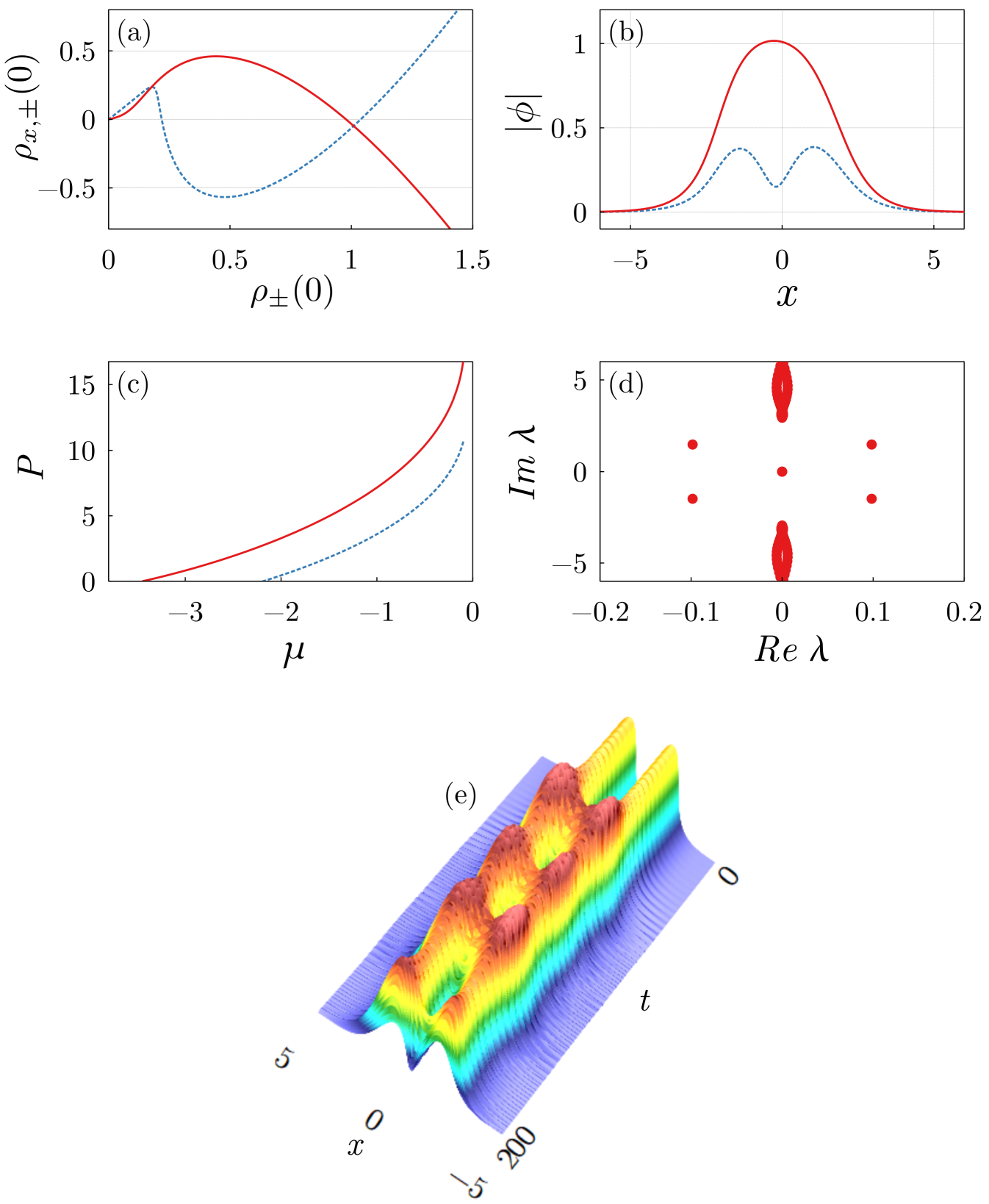}%
	\end{center}
	\caption{(a) Dependencies $\rho_+(0)$ vs.  $\rho_{+,x}(0)$ (solid red curve) and $\rho_-(0)$ vs.  $\rho_{-,x}(0)$ (dotted blue curve) obtained by the numerical shooting procedure for $\mu=-2$.  (b)  Amplitudes of nonlinear modes $|\psi(x)| = |\phi(x)|$ corresponding to the two  intersections in (a). Solid red and dotted blue profiles correspond to the fundamental and first excited states, respectively.  (c) Continuous curves corresponding to the families of fundamental (solid red curve) and first excited (dotted blue curve) nonlinear modes. In (a)-(c) we use $F(|\psi|^2)=0$ and $\sigma=1$. (d) Linear stability eigenvalues $\lambda$ (shown with red circles) for the mode from the excited  family  at $\mu=-3$ for $F(|\psi|^2) = -|\psi|^2$ and   $\sigma=0.25$. (e) Temporal evolution $|\Psi(x,t)|$ of an initial condition taken as a slightly perturbed stationary  mode  whose  linear-instability spectrum is shown in (d).}
	\label{fig:asym}
\end{figure}

Proceeding now to a numerical illustration, we look for stationary modes of equation   (\ref{eq:dnls}) with the derivative nonlinearity  and therefore consider $A(x,\rho^2(x)) = w(x) - \sigma\rho^2(x)/2$ with an asymmetric base  function $w(x)$ in the form
\begin{equation}
\label{eq:w}
w(x) = \tanh2(x-2.5)  - \tanh(x-2.5).
\end{equation}
For linear Wadati potentials, this  base function  has been considered in \cite{quartets}.  To keep the model as simple as possible but still nontrivial, in this example   we  remove the conventional  nonlinearity by setting $F \equiv 0$. In Fig.~\ref{fig:asym}(a) we illustrate the numerical shooting procedure by presenting the dependencies    $\rho_+(0)$ \emph{vs.}  $\rho_{+,x}(0)$   and $\rho_-(0)$ \emph{vs.}  $\rho_{-,x}(0)$ obtained from the numerical solution of the system (\ref{eq:hydro1})--(\ref{eq:hydro2}) under the gradual increase  of shooting parameters $C_+$ and $C_-$ departing from zero, and  for fixed value of $\mu$. Apart from the origin (which corresponds to the trivial zero solution with $C_+=C_-=0$), the shown curves feature two intersections, which correspond to the particular values of the  shooting parameters   $(C_+, C_-)$   for which the first two equations of system (\ref{eq:3eqs}) are satisfied. It has been checked  that the third equation of this system is also satisfied for both intersections and hence  they   indeed correspond to two localized modes. Spatial profiles $|\psi(x)|=|\phi(x)|$ of these modes are shown in Fig.~\ref{fig:asym}(b). Gradually changing $\mu$  and tracing the found intersections, we construct two continuous families which are visualized in Fig.~\ref{fig:asym}(c) in the form of continuous dependencies $P(\mu)$, where $P=\int_{-\infty}^{\infty} |\psi|^2dx$ is the squared $L^2$-norm of the solution (that corresponds to the optical beam power). In the limit $P\to 0$ the solutions become small-amplitude, and the corresponding values of $\mu$ approach    the eigenvalues of the underlying linear eigenvalue problem which can be formally obtained from (\ref{eq:old}) by setting $F(|\psi|^2)=0$.

While the discussion in this subsection has been limited to the spatially localized modes whose amplitude $|\rho(x)|$ vanishes rapidly  enough as $x\to \infty$ and $x\to-\infty$, the consideration can be extended toward inclusion of stationary modes with nonvanishing asymptotic behavior. In particular, similar arguments can be elaborated for kink-like solutions with $\lim_{x\to\pm \infty} \rho(x) = r_{\pm}$, where $r_+$ and $r_-$ are nonzero constants, provided that the corresponding asymptotic behavior can be exhaustively  described using several shooting parameters.

\subsection{Eigenvalue quartets in the linear-instability spectrum} 

For linear Wadati potentials, it has been found numerically that the complex eigenvalues  of the linear-stability operator always appear in quartets $(\lambda, \lambda^*, -\lambda, -\lambda^*)$ \cite{quartets}. This numerical observation (which, to the best of our knowledge,   has not yet received any analytical confirmation)  is rather intriguing,  because the eigenvalue quartets are more typical to Hamiltonian systems or to setups constrained by  some explicit  symmetry (such as $\PT$-symmetric systems).  In the meantime, the GNLSE with Wadati potentials does not apparently feature a Hamiltonian structure and has no readily evident symmetry.  In order to examine if the eigenvalue quartets persist  in our generalized model,   we perform the standard linearization procedure   using the perturbed stationary-mode substitution in the form $\Psi(x,t) = e^{-i\mu t} [\psi(x) + a(x)e^{\lambda t} + b^*(x)e^{\lambda^* t}]$, where $a(x)$ and $b(x)$ are small perturbations, and complex $\lambda$ is linear stability eigenvalue. Positive real part of $\lambda$ defines the increment of an eventual instability. Using this substitution in Eq.~(\ref{eq:dnls}) and keeping only the linear in $a(x)$ and $b(x)$ terms, we obtain a linear-stability  eigenproblem for eigenvalue $\lambda$ and eigenvector $(a, b)^T$ (since the linearization procedure is straightforward, we do not present the corresponding linear stability equations herein). For a numerical illustration we consider  Eq.~(\ref{eq:dnls}) with  the the base function given by (\ref{eq:w}) and cubic nonlinearity $F(|\psi|^2) = -|\psi|^2$. Then for $\sigma=0$ our model recovers that from \cite{quartets}, where the existence of quartets has been observed numerically. Increasing $\sigma$ and evaluating numerically the linear-stability spectra, we observe that the eigenvalue quartets persist  for $\sigma\ne 0$.  For example, for $\mu=-3$ and $\sigma=0.25$ we obtain eigenvalues $\lambda_{1,2} \approx 0.0983725 \pm i 1.484634$ and  $\lambda_{3,4}  \approx  -0.0983725 \pm i 1.484634$, i.e., the numerical results allow  to conjecture the found eigenvalues indeed form a quartet, at least with the accuracy $~10^{-6}$. A rigorous proof of this fact remains an interesting issue for future studies.

To corroborate the presence of the instability, we have simulated nonlinear dynamics governed by Eq.~(\ref{eq:dnls})  using a modification of the finite-difference Crank-Nicolson scheme adapted for the derivative nonlinearity \cite{LiLiShi,GuoFang}. Choosing the slightly perturbed unstable stationary  mode as an initial condition, from its evolution shown in Fig.~\ref{fig:asym}(e) we observe that the  stationary mode  preserves its shape for $t\lesssim 50$, but for larger times the shape of  numerical solution oscillates aperiodically.

\subsection{Symmetry-breaking bifurcation and non-$\PT$-symmetric modes in the $\PT$-symmetric case}

\begin{figure}
	\begin{center}
		\includegraphics[width=0.99\columnwidth]{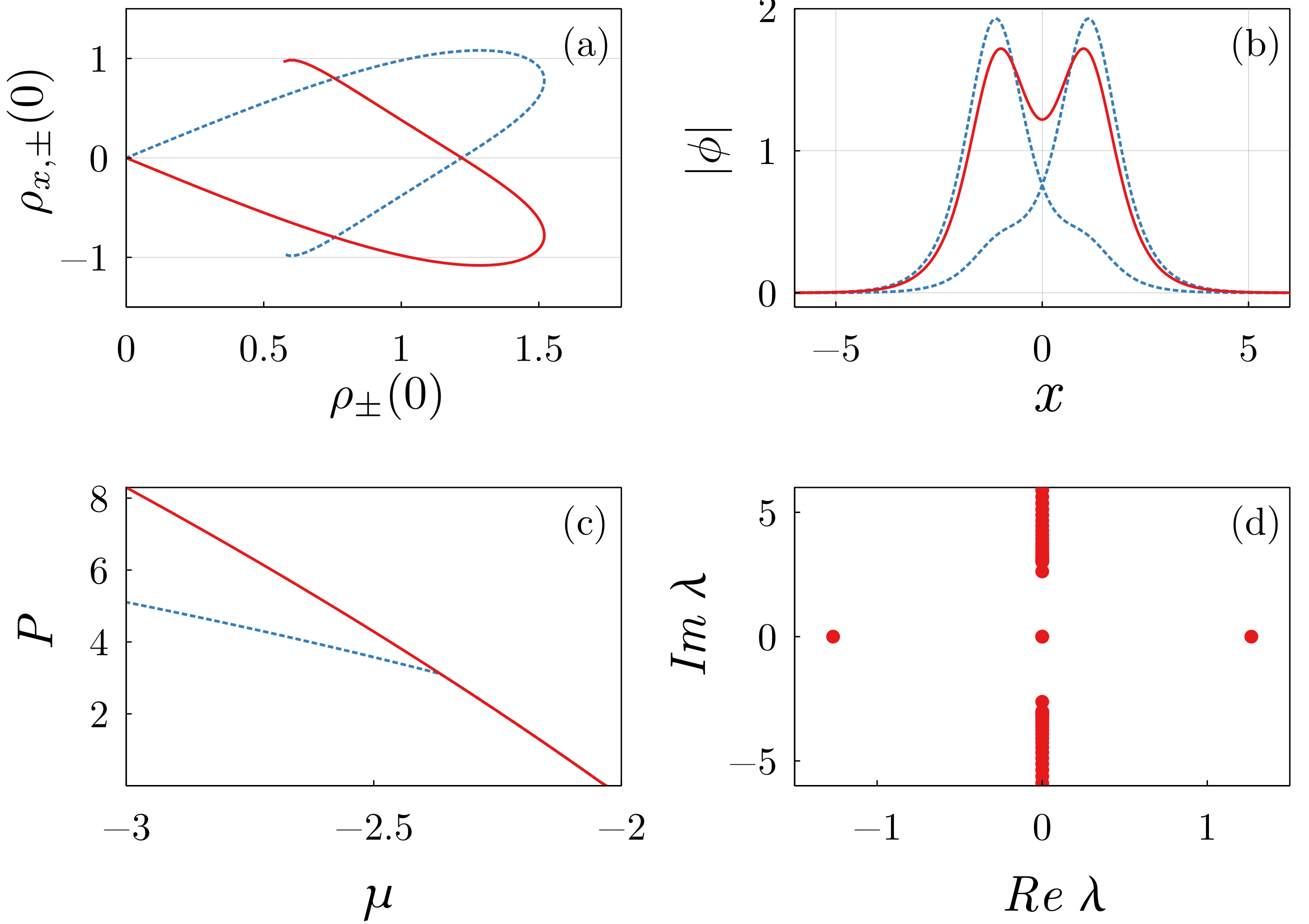}%
	\end{center}
	\caption{(a) Dependencies $\rho_+(0)$ vs.  $\rho_{+,x}(0)$ (solid red curve) and $\rho_-(0)$ vs.  $\rho_{-,x}(0)$ (dotted blue curve) obtained by the numerical shooting procedure for $\PT$-symmetric system at $\mu=-3$ constructed by increasing the shooting parameters from $C_\pm=0$ to $C_\pm=34$.    (b)  Amplitudes of nonlinear modes corresponding to the intersections in (a). Solid red  curve corresponds to the $\PT$-symmetric soliton and  two dotted  blue  profiles correspond to a pair of symmetry-broken solitons.   (c) Continuous families of $\PT$-symmetric (solid red curve) and asymmetric  (dotted blue curve) solitons.   (d) Linear stability eigenvalues for the $\PT$-symmetric soliton at $\mu=-3$. In this figure  $F(|\psi|^2)=-|\psi|^2$ and $\sigma=0.4$.}
	\label{fig:PT}
\end{figure}

\begin{figure}
	\begin{center}
		\includegraphics[width=0.99\columnwidth]{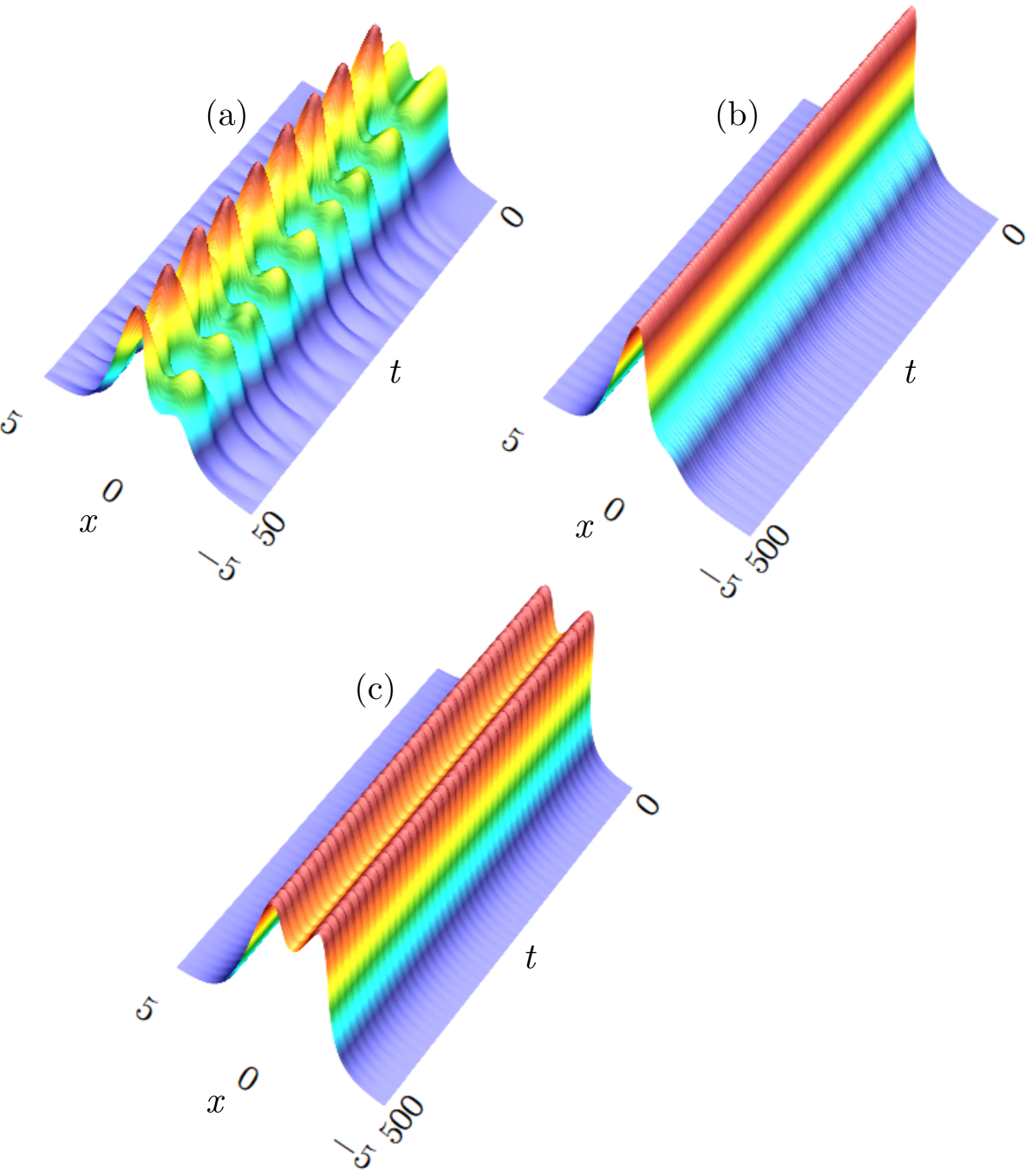}%
	\end{center}
	\caption{Temporal evolutions of initial conditions taken as slightly perturbed stationary modes from Fig.~\ref{fig:PT}: unstable symmetric (a) and stable asymmetric (b) modes above the symmetry-breaking bifurcation (specifically,   at $\mu=-3$) and stable symmetric mode below the symmetry-breaking bifurcation (specifically,  at $\mu=-2.2$). The plots show $|\Psi(x,t)|$.}
	\label{fig:dynPT}
\end{figure}

Next, we illustrate that in the case when the introduced  generalized model becomes $\PT$ symmetric, it can support asymmetric (more precisely, not $\PT$ symmetric) nonlinear modes. As in the case of linear Wadati potentials, Eq.~(\ref{eq:dnls}) becomes  $\PT$ symmetric \cite{Bender}, i.e., invariant under the transformation $x\to -x$, $t\to -t$ and $i\to -i$,  when   the base function $w(x)$ is even. Symmetric and  asymmetric modes can be searched using the   numerical shooting approach outlined above.  For a numerical illustration, we use a bimodal base function in the form $w(x) = 2e^{-(x-1.2)^2} + 2e^{-(x+1.2)^2}$ which results in an effectively double-well potential. In Fig.~\ref{fig:PT}(a) we plot  the corresponding  dependencies    $\rho_+(0)$ \emph{vs.}  $\rho_{+,x}(0)$   and $\rho_-(0)$  \emph{vs.}  $\rho_{-,x}(0)$. Three  intersections in Fig.~\ref{fig:PT}(a) correspond to a pair of asymmetric modes (the two intersections with   $\rho_x(0)\ne 0$) and to a $\PT$-symmetric mode (the intersection   with $\rho_x(0)=0$). Spatial profiles of symmetric and asymmetric modes are plotted in   Fig.~\ref{fig:PT}(b). Changing the value of $\mu$, we construct continuous families of $\PT$-symmetric and asymmetric modes.  Two  asymmetric families bifurcate from the   symmetric one after a supercritical pitchfork bifurcation: in other words, the symmetric family is stable for small $P$, but becomes unstable right after the bifurcation.    A representative plot of  linear stability eigenvalues  for the symmetric family past the symmetry-breaking bifurcation is plotted in Fig.~\ref{fig:PT}(d). In Fig.~\ref{fig:dynPT} we present several representative examples of numerically computed nonlinear dynamics in the $\PT$-symmetric system. In agreement with the predictions of the  linear-stability analysis, we observe that above the symmetry-breaking bifurcation the symmetric family is unstable [Fig.~\ref{fig:dynPT}(a)]: the initially perturbed symmetric stationary mode shortly evolves to a breather-like oscillating entity. The asymmetric families [Fig.~\ref{fig:dynPT}(b)] and the symmetric family below the symmetry-breaking bifurcation are stable, and the corresponding numerical solutions preserve the stationary shape for indefinitely long time.

\subsection{Absence of the dynamical conservation law}

Results of the previous subsections do not imply that  nonlinear Wadati potentials simply inherit any property  of linear Wadati potentials without even a minor modification. As an example of a dissimilarity   between the two types of potentials, we recall that   linear Wadati potentials admit a dynamical (i.e., $t$-independent) conserved quantity given as \cite{integral}
\begin{equation}
\label{eq:J}
J  = \int_{-\infty}^\infty [i\Psi \Psi_x^* - w(x) |\Psi|^2]dx.
\end{equation}
Indeed, for linear Wadati potentials from Eqs.~(\ref{eq:intro})--(\ref{eq:Wadati}) we compute $dJ/dt = 0$. In the meantime, for nonlinear Wadati potentials from Eq.~(\ref{eq:dnls}) we evaluate
\begin{eqnarray*}
\frac{dJ}{dt} = \sigma \int_{-\infty}^\infty \left[ 2w  (|\Psi|^4)_x dx + i   (|\Psi|^2)_x (\Psi_x \Psi^* - \Psi_x^* \Psi)\right] dx\nonumber\\[2mm] = \sigma   \int_{-\infty}^\infty    (|\Psi|^4)_x [2w - (\arg \Psi)_x]dx.
\end{eqnarray*}
Hence the dynamical conserved quantity (\ref{eq:J}) does not carry over  to  nonlinear Wadati potentials. Of course, this does not imply that the extended model (\ref{eq:dnls}) does not have   any dynamical conservation law at all. However, so far we have not been able to find a generalization of the dynamical integral (\ref{eq:J}) for  nonlinear Wadati potentials.

\section{Conclusion}
\label{sec:concl}

Nonlinear Schr\"odinger equation with  complex Wadati-type  potentials is   known to feature a variety of unusual and intriguing properties. In the meantime, the    generalizations of Wadati potentials are rather scarce, and  most of the  activity in this direction is presently  limited by spatially homogeneous power-law and saturating nonlinearities. In this paper, we have  proposed a significant  extension of  Wadati potentials. The main idea of our approach is to consider    the base function of a Wadati potential as depending not only on  the spatial coordinate but also on the amplitude of the field.   The resulting  extended  model   admits a conserved (i.e., independent of the transverse coordinate) quantity.  Using this   conserved quantity, we have employed a demonstrative computation approach to argue that the  generalized model supports continuous families of bright solitons. The numerical study of the linear-instability spectra   indicates  that unstable solitons feature eigenvalue quartets, which  is another remarkable peculiarity of Wadati potentials. We have demonstrated that the generalized parity-time-symmetric model supports a supercritical symmetry-breaking bifurcation that gives birth to continuous families of non-$\PT$-symmetric solitons. Numerical simulations of nonlinear dynamics indicate that unstable stationary modes can dynamically transform to  nearly periodic breather-like solutions.   The introduced model also admits a straightforward generalization of constant-amplitude nonlinear waves known to exist in conventional Wadati potentials.  In contrast to the previously known models with Wadati potentials, our generalization incorporates spatially modulated nonlinearity, where the shape of the modulation is determined by the base function of the Wadati potential. Additional nonlinear terms can  include spatially-uniform   nonlinear dispersion  or   derivative nonlinearity.  Our findings essentially broaden the class of systems which enjoy the unique combination of peculiar features of nonlinear and non-Hermitian systems    typical to Wadati-type and $\PT$-symmetric potentials.

\section*{Acknowledgments}
The work was supported by the Priority 2030 Federal Academic Leadership Program.

\end{document}